\documentclass{article}
\usepackage{spconf,amsmath,graphicx}
\usepackage[table,xcdraw]{xcolor}
\usepackage{adjustbox}
\usepackage{url}

\title{MULTI-LEVEL BATCH NORMALIZATION IN DEEP NETWORKS FOR INVASIVE DUCTAL CARCINOMA CELL DISCRIMINATION IN HISTOPATHOLOGY IMAGES}
\name{Francisco Perdig\'on Romero$^{\star }$ \qquad An Tang$^{\dagger}$ \qquad Samuel Kadoury$^{\star\dagger}$ }
\address{$^{\star}$MedICAL Laboratory, Polytechnique Montreal, Montr\'eal, Canada\\
	$^{\dagger}$Centre de recherche du CHUM (CRCHUM), Montr\'eal, Canada}

\begin{document}
%
\maketitle

\begin{abstract}
Breast cancer is the most diagnosed cancer and the most predominant cause of death in women worldwide. Imaging techniques such as the breast cancer pathology helps in the diagnosis and monitoring of the disease. However identification of malignant cells can be challenging given the high heterogeneity in tissue absorbotion from staining agents. In this work, we present a novel approach for Invasive Ductal Carcinoma (IDC) cells discrimination in histopathology slides. We propose a model derived from the Inception architecture, proposing a multi-level batch normalization module between each convolutional steps. This module was used as a base block for the feature extraction in a CNN architecture. We used the open IDC  dataset in which we obtained a balanced accuracy of 0.89 and an F1 score of 0.90, thus surpassing recent state of the art classification algorithms tested on this public dataset.

\end{abstract}
\begin{keywords}
breast cancer, pathology, deep learning, batch normalization, convolutional neural networks

\end{keywords}
\section{Introduction}
\label{sec:intro}
According to a recent census from 140 countries worldwide \cite{WorldCancerReport}, breast cancer is the most frequently diagnosed cancer in women, with an estimated incidence of 1.7 million new cases per year. It represents 25\% of all types of cancers diagnosed in women. Furthermore, 15\% of cancer deaths in women are due to breast cancer \cite{CancerIncidence}. In recent years, a combination of improved detection techniques and earlier diagnosis, added to more effective treatments, has caused the mortality rate to decrease in developed or emerging countries \cite{WorldCancerReport}.

This has motivated groups to pursue efforts in improving detection methods at early stages of the disease. Breast cancer detection through histopathology images allows to quantify the cancer cell ratio with respect to healthy cells. Oncologists use this information to determine cancer aggressiveness and select the appropriate treatment option. Previously, researchers have worked on methods for cancer/healthy cell classification \cite{SOTA1,SOTA2}. However, the scientific community is still working to improve these classification methods by identifying predominant features in cell images.

Convolutional neuronal networks (CNN) have become increasingly popular in the processing of histopathological images \cite{DLSurvey}. The main difference with respect to traditional machine learning approaches is that feature extraction is done intrinsically and hierarchically. This allows the network to choose in a an unsupervised fashion the features for optimal cell discrimination. In this work, we present a new CNN-based model for cancer/healthy cell classification on histopathology slides (HPS). We leverage these recent discoveries in deep learning, particularly using the Inception architecture \cite{InceptionV1} by combining regularization techniques such as batch normalization \cite{BatchNorm_ref} in order to reduce the intrinsic covariate shift.


\section{Methods and materials}
\label{sec:format}

\subsection{Dataset}

\begin{figure}[htb]
\begin{minipage}[b]{1.0\linewidth}
  \centering
  \centerline{\includegraphics[width=8.5cm]{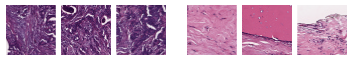}}
\end{minipage}

\caption{Left: three IDC cell patches; Right: three normal tissue patches.}
\label{fig:Dataset}
\end{figure}

We used a publicly available dataset \footnote{Available at \url{http://andrewjanowczyk.com/wp-static/IDC_regular_ps50_idx5.zip}} that was first introduced by Cruz-Roa et al. \cite{SOTA1} for Invasive Ductal Carcinoma (IDC) identification. The histopathological images of the dataset were not released in their original format but in RGB patches of 50x50 pixels. The zoom factor  was 2.5x (4μm/pixel). The dataset contained 277,525 patches from 279 subjects. The patches were comprised of  28.39\% invasive ductal carcinoma cells, while the remaining patches consisted of healthy tissue or non-invasive ductal carcinoma (see Figure \ref{fig:Dataset}).

\subsection{Proposed model}
The Inception module \cite{InceptionV1} introduced by Szegedy C. et al. was initially proposed to deal with the size variability problem in hierarchical representations. The presented solution was simple and achieved excellent results by using parallel filters with different kernel sizes, and allowed the back-propagation process to select which filter would be used in each case.

\begin{figure}[htb]
\begin{minipage}[b]{1.0\linewidth}
  \centering
  \centerline{\includegraphics[width=8.5cm]{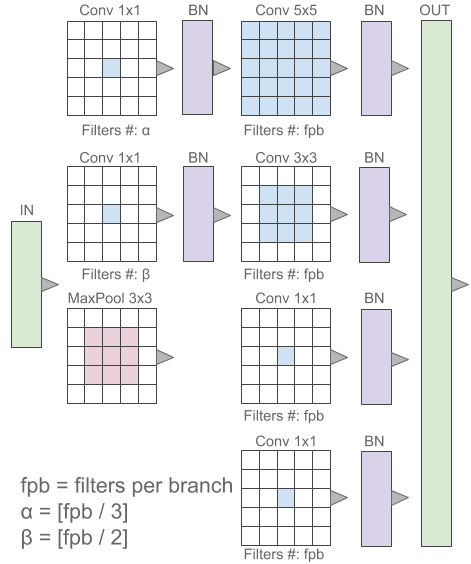}}
\end{minipage}

\caption{Schematic illustration of the proposed Inception module.}
\label{fig:Arch}
\end{figure}

Our model is composed of blocks derived from Inception that integrates batch normalization (BN) \cite{BatchNorm_ref} after each convolution step. For each Inception module, the feature number per branch is determined during validation. Due to the concatenation that takes place in each module, the feature output number is four times greater than the feature number per branch value. In the branches that contain the filters with 5x5 and 3x3 kernels, dimensionality reduction is applied to minimize memory consumption through filters with 1x1 kernels. The number of dimensionality reduction filters is set by the ($\alpha$) and ($\beta$) parameters, which vary depending on the feature number per branch (see Figure \ref{fig:Arch}).

The batch normalization steps reduce the internal covariate shift, improving network training stability. This aids in the training as the technique re-parametrizes the underlying optimization problem thereby making the gradients more predictable \cite{newBNRef}.

MaxPool layers are used to reduce the size of the internal network representations. These layers increase the number of extracted features without excessive memory consumption and provide a certain level of rotational invariance. In the final stages where the classification is performed, we include two fully connected layers of 512 neurons with batch normalization and dropout of 0.4 each. The network output contains two neurons followed by a SoftMax layer. See Table \ref{tab:Architecture} for the detailed architecture.

\begin{table}[htb]
\caption{Details of the framework architecture.} \label{tab:Architecture}
\begin{tabular}{ccc}
Block Type         & Block Name          & Feature \# \\ \hline
\rowcolor[HTML]{EFEFEF} 
Feature Extraction & Inception           & 64              \\
Feature Extraction & Inception           & 64              \\
\rowcolor[HTML]{EFEFEF} 
Feature Extraction & MaxPool             & -               \\
Feature Extraction & Inception           & 128             \\
\rowcolor[HTML]{EFEFEF} 
Feature Extraction & Inception           & 126             \\
Feature Extraction & MaxPool             & -               \\
\rowcolor[HTML]{EFEFEF} 
Feature Extraction & Inception           & 256             \\
Feature Extraction & Inception           & 256             \\
\rowcolor[HTML]{EFEFEF} 
Feature Extraction & MaxPool             & -               \\
Feature Extraction & Inception           & 512             \\
\rowcolor[HTML]{EFEFEF} 
Feature Extraction & Inception           & 512             \\
Classification     & Fully connected     & 256             \\
\rowcolor[HTML]{EFEFEF} 
Classification     & Batch Normalization & -               \\
Classification     & Dropout (0.4)       & -               \\
\rowcolor[HTML]{EFEFEF} 
Classification     & Fully connected     & 256             \\
Classification     & Batch Normalization & -               \\
\rowcolor[HTML]{EFEFEF} 
Classification     & Dropout (0.4)       & -               \\
Classification     & Fully connected     & 2               \\
\rowcolor[HTML]{EFEFEF} 
Classification     & Softmax             & -              
\end{tabular}
\end{table}

\subsection{Training Protocol}

In this section, we present the training protocol with the parameter selection strategies.

\textbf{Padding}: Most of the patch images were 50x50 pixels in size, but in some isolated cases, images which contained 1-2 fewer rows or columns were zero-padded.

\textbf{Normalization}: Mean centering and standard deviation normalization was applied to each image. This decreases the variability of the input data, thus improving the training stability.

\textbf{Train/Validation/Test split}: We train the network with a balanced training set of 94,543 images. The validation set contained 31,514 images, also balanced. For the final evaluation, our test set is unbalanced, the total number of images were 151,465 images, of these 15,757 images contain cancer cells.

\textbf{Batch size}: A batch size of 32 images was used for training. This makes the training faster as it takes advantage of the speed-up of matrix-matrix products \cite{bengio2012practical}.

\textbf{Optimization and Learning Rate Scheduling}: Adam optimization algorithm \cite{Adam} was used with an initial learning rate of 10$^{-3}$. This learning rate was reduced by a factor of 2 after 10 epochs without improvements in the validation set accuracy. The allowed minimum learning rate was 10$^{-10}$.

\textbf{Iterations}: The initial number of epochs was 10$^{3}$, but due to the early stopping criterion that was implemented, the final number of epochs was 55. For early stopping, the validation set accuracy was monitored, and stopped after 50 epochs if no improvements was detected.

Training time on an NVIDIA Titan Xp GPU was 170 minutes, while the inference time per single patch was 3 ms (89 ms on an I7 7th generation CPU). The above timings were calculated for the proposed model, include the Inception architecture.


\section{Results and Discussion}

Table \ref{tab:Results} shows a comparison between the balanced accuracy and F1-score obtained with our method and previous methods using the same dataset. Both metrics are designed to deal with unbalanced datasets because in these cases, metrics such as accuracy can lead to misinterpretations in the classifier efficiency. Our model showed an improvement in the results obtained by Cruz-Roa et al. \cite{SOTA1}, that presented a custom deep neural network, and the results obtained by Janowczyk and Madabhushi \cite{SOTA2} who used AlexNet architecture \cite{AlexNet} for the classification task in the same dataset.

\begin{table}[htb]
\begin{center}
\caption{Ensemble evaluation results comparing our model with others approaches in the same dataset.} \label{tab:Results}

\begin{tabular}{l|cc}
\textbf{Model}                 & \textbf{\begin{tabular}[c]{@{}c@{}}Balanced \\ Accuracy\end{tabular}} & \textbf{F1 score} \\ \hline
2D-CNN \cite{SOTA1}       & 0.842                                                                 & 0.718             \\
AlexNet with BN \cite{SOTA2} & 0.847                                                                 & 0.765             \\
Original InceptionNet \cite{InceptionV1}       & 0.868                                                       & 0.883  \\
\textbf{Proposed method}              & \textbf{0.890}                                                        & \textbf{0.897}   
\end{tabular}
\end{center}
\end{table}

\begin{figure}[htb]
\begin{minipage}[b]{1.0\linewidth}
  \centering
  \centerline{\includegraphics[width=8.5cm]{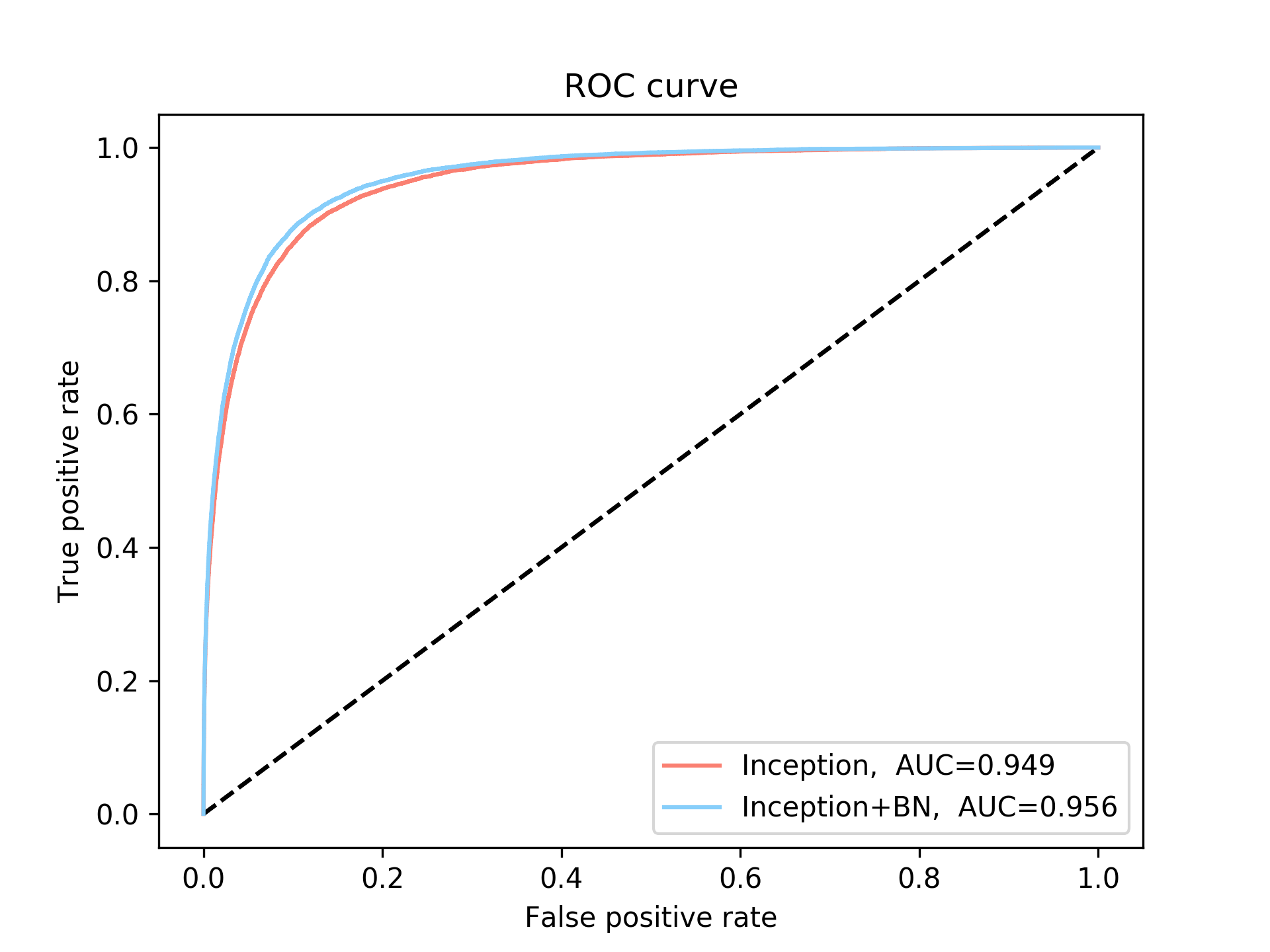}}
\end{minipage}

\caption{Models performance represented by the ROC curves.}
\label{fig:ROC}

\begin{minipage}[b]{1.0\linewidth}
  \centering
  \centerline{\includegraphics[width=8.5cm]{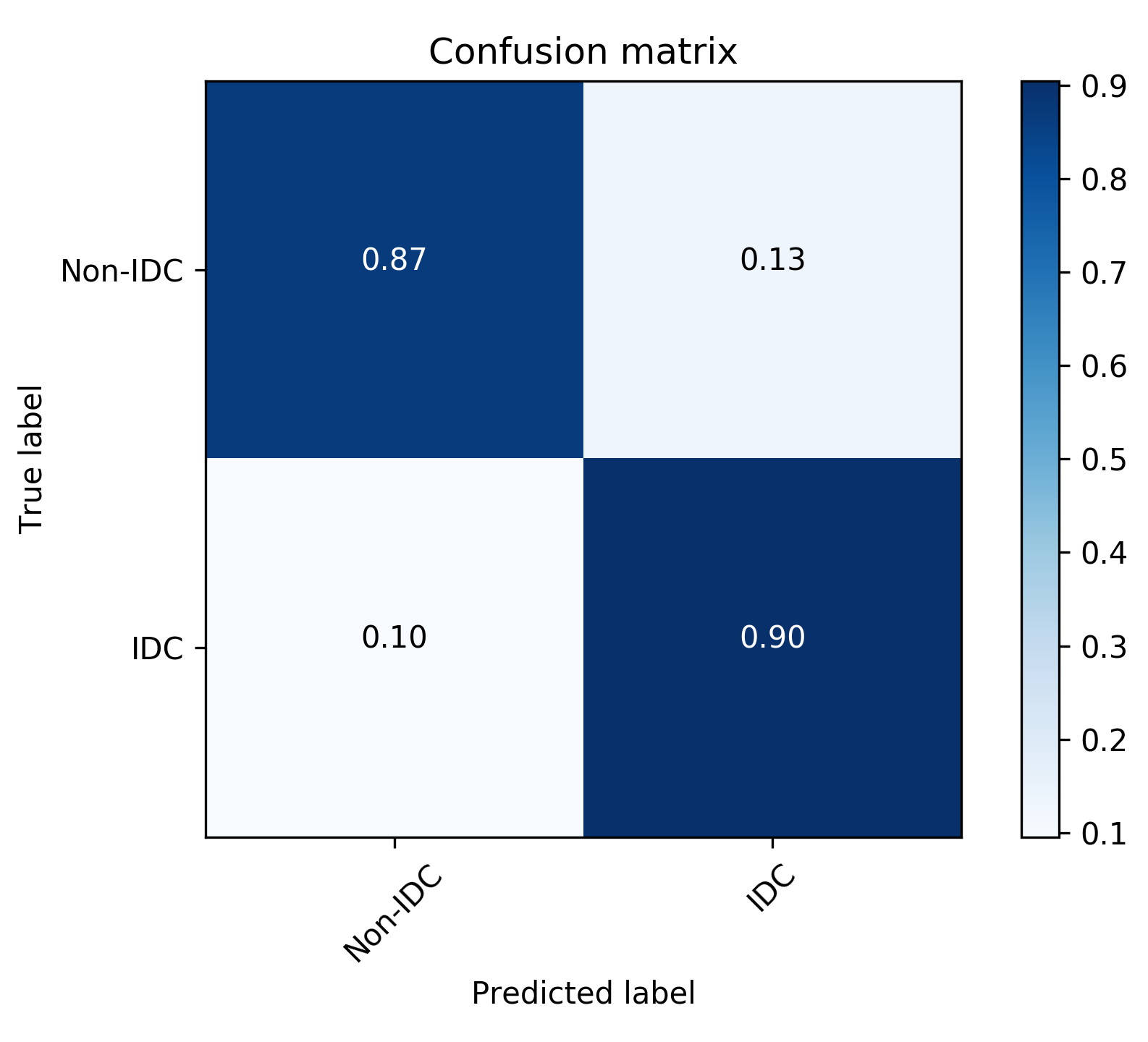}}
\end{minipage}

\caption{Model performance evaluated by the confusion matrix.}
\label{fig:CM}
\end{figure}

Figures \ref{fig:ROC} and \ref{fig:CM} shows the efficiency of our classifier. The area under the curve (AUC) of the receiver operating characteristic (ROC) was 0.956 for Inception+BN, while the original Inception module was 0.949. The improvement in accuracy was 2.2\% compared to the original model. Figure \ref{fig:CM} shows the confusion matrix of the obtained results for the proposed architecture.

These results demonstrate the effectiveness of our Inception module variant. Due to network's  multi-scale capability of processing information through filters with different kernel sizes, the network is able to extract highly discriminative features. Furthermore, the use of batch normalization aids the training by reducing internal covariate shift, thus increasing the stability of the deep neural network. Finally, the training strategy allowed us to obtain the maximum benefit of each learning rate step, improving the convergence of the network.

\begin{figure*}[htb]
    \begin{center}
        \begin{tabular}{cc}
            \includegraphics[height=6.4cm]{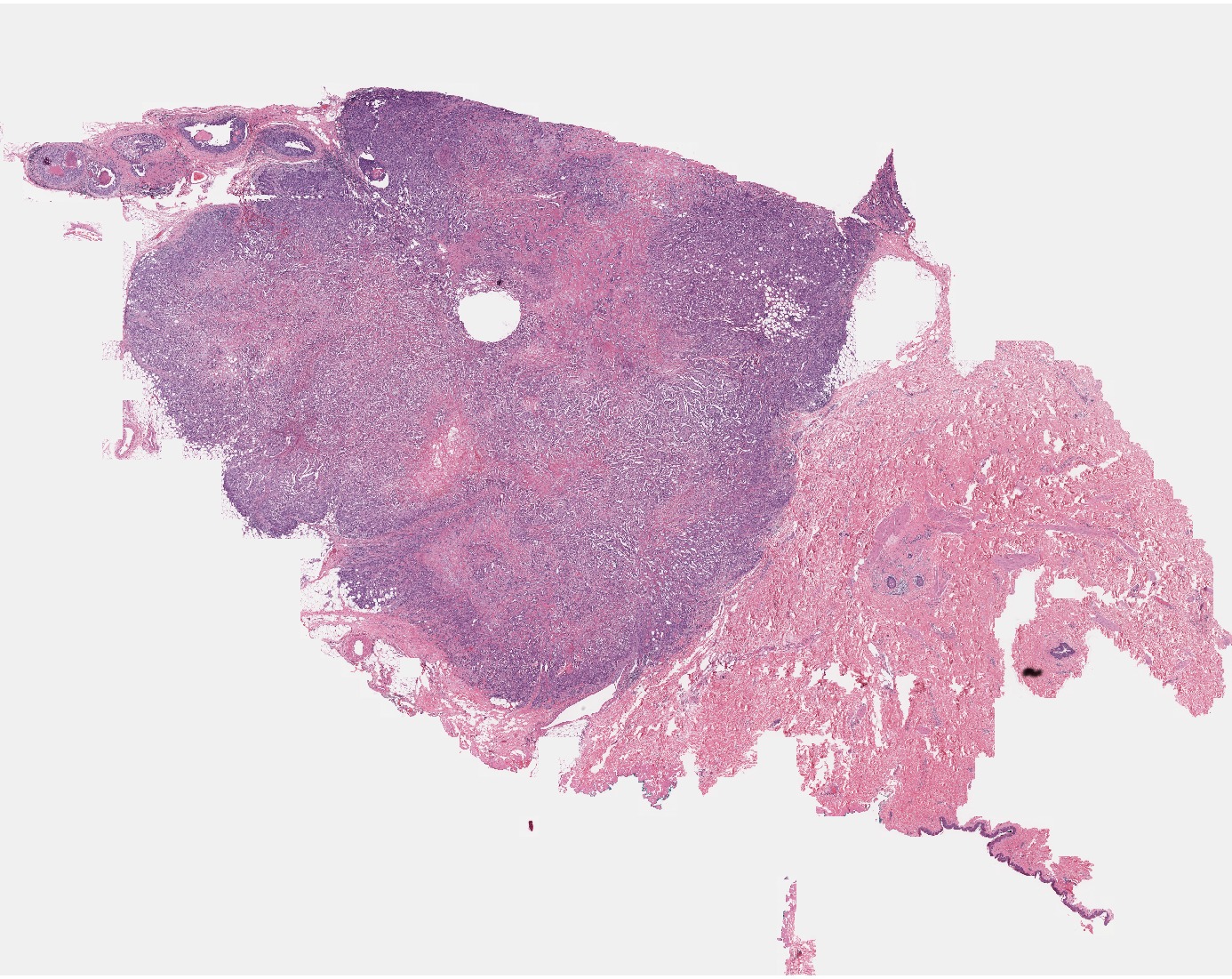} & \includegraphics[height=6.4cm]{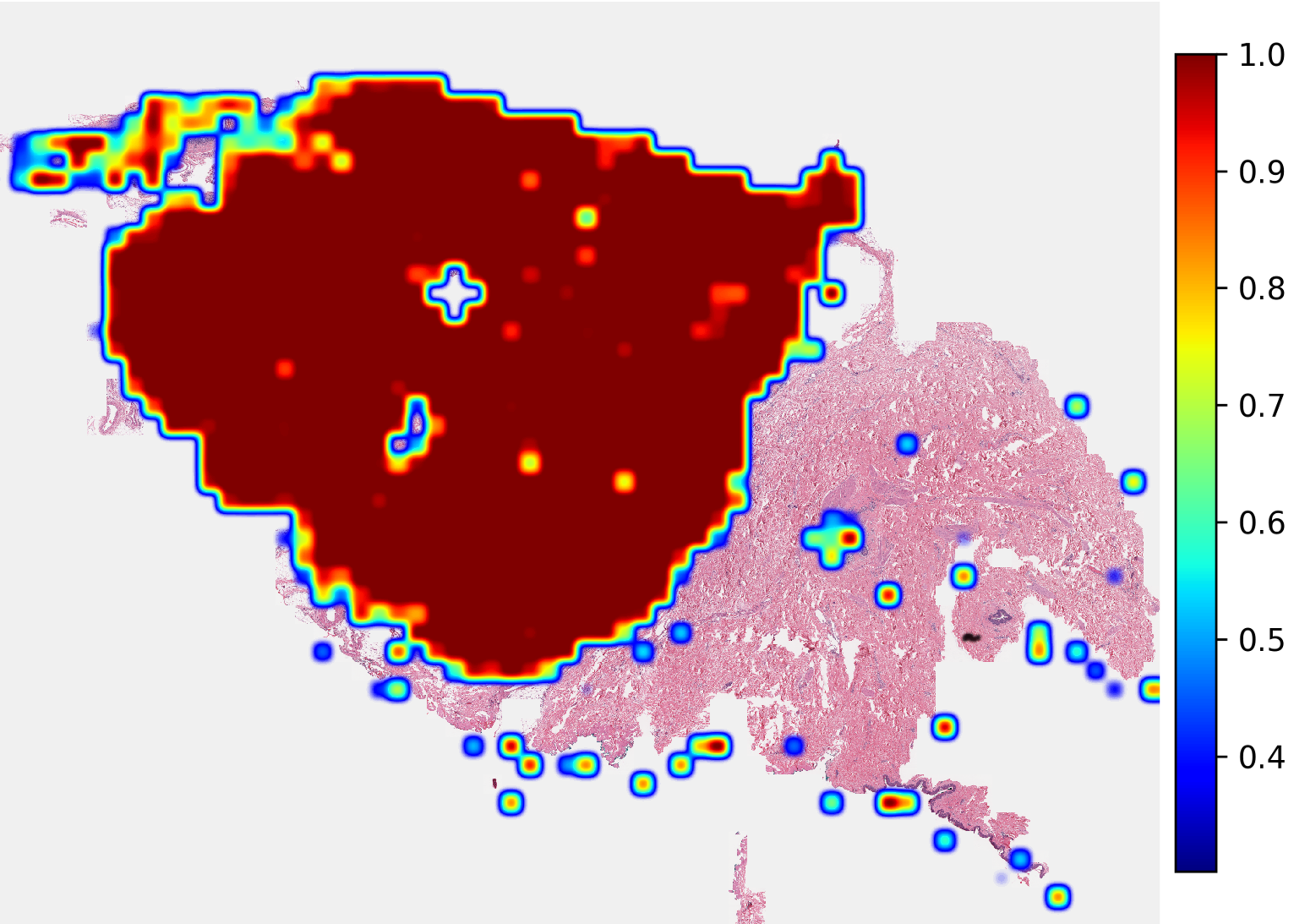}\\
            \small{(a)} & \small{(b)}
        \end{tabular}
    \end{center}
    \vspace{-0.5cm}
    \caption{IDC cell prediction results. (a) Original reconstructed pathology slide. (b) Prediction heat map overlapped to the original slide.}
    \label{fig:Predictions}
\end{figure*}

Because the original patches in the database contain patch positions, this enables to reconstruct the entire original slides. The trained models were then applied at the patch level to generate entire processed slides. Because our model processes patches of 50x50 pixels, the generated heatmap appeared ``pixelated''. To improve the final rendering, we applied a Gaussian filter with a 25x25 kernel. Figure \ref{fig:Predictions} (a) shows the slide number 14154, while Figure \ref{fig:Predictions} (b) shows an overlap of the predictions heat map generated by our model. Our framework can be applied to HPS of other datasets, provided they have a similar zoom factor of x2.5.

\section{Conclusion}
\label{sec:typestyle}

We presented a novel approach for Invasive Ductal Carcinoma cell discrimination in histopathology slides. The model is composed of Inception variant blocks with combined  batch normalization steps to reduce internal covariances. Results surpass state of the art classification methods which were trained and tested on the same dataset. The quantitative evaluation of the results was carried out through balanced accuracy and F1-score, metrics used for unbalanced datasets. Future work will be focused on the inclusion of residual blocks \cite{ResNet_ref} to deal with the vanishing gradients during the training, which in theory will lead to improved accuracy.

\bibliographystyle{IEEEbib}
\bibliography{refs}

\end{document}